# Structural color tuning in 1D photonic crystals with electric field and magnetic field

Eduardo Aluicio-Sarduy[a,b], Simone Callegari[b], Diana Gisell Figueroa del Valle[a,c], Andrea Desii[a], Ilka Kriegel[d], Francesco Scotognella[a,c,]*

[a]Center for Nano Science and Technology@PoliMi, Istituto Italiano di Tecnologia, Via Giovanni Pascoli, 70/3, 20133, Milan, Italy; [b]Dipartimento di Chimica, Materiali e Ingegneria Chimica "Giulio Natta", Politecnico di Milano, Piazza Leonardo da Vinci 32, 20133 Milano, Italy; [c]Dipartimento di Fisica, Politecnico di Milano, Piazza Leonardo da Vinci 32, 20133 Milano, Italy; [d] Department of Nanochemistry, Istituto Italiano di Tecnologia (IIT), via Morego, 30, 16163 Genova, Genova, Italy
*francesco.scotognella@polimi.it; phone 39 02 2399-6056; fax 39 02 2399-6126; www.fisi.polimi.it/en/people/scotognella

## ABSTRACT

A tuning of the light transmission properties of 1D photonic structures employing an external stimulus is very attracting and opens the way to the fabrication of optical switches for colour manipulation in sensing, lighting, and display technology. We present the electric field-induced tuning of the light transmission in a photonic crystal device, made by alternating layers of silver nanoparticles and titanium dioxide nanoparticles. We show a shift of around 10 nm for an applied voltage of 10 V. We ascribe the shift to an accumulation of charges at the silver/$TiO_2$ interface due to electric field, resulting in an increase of the number of charges contributing to the plasma frequency in silver, giving rise to a blue shift of the silver plasmon band, with concomitant blue shift of the photonic band gap. The employment of a relatively low applied voltage gives the possibility to build a compact and low-cost device [1]. We also propose the fabrication of 1D photonic crystal and microcavities employing a magneto-optical material as TGG ($Tb_3Ga_5O_{12}$). With these structures we can observe a shift of 22 nm with a magnetic field of 5 T, at low temperature (8 K). The option to tune the colour of a photonic crystal with magnetic field is interesting because of the possibility to realize contactless optical switches [2]. We also discuss the possibility to achieve the tuning of the photonic band gap with UV light in photonic crystals made with indium tin oxide (ITO).

## 1. INTRODUCTION

Photonic crystals are peculiar materials in which the alternation of materials with different refractive indexes occur at a length scale comparable with the wavelength of light [3–7]. A characteristic of these materials is the structural colour, due to the photonic band gap that is an energy region in which photons are not allowed to propagate through them. Thus, the structural colour is not due to the absorption of a dye or colour centers, but it is strictly connected to the mesoscopic structure of the material. The simplest photonic crystal is the one-dimensional case, where two materials are alternated in one direction, as for example in a multilayer structure. The center of the photonic band gap follows the Bragg-Snell law

$$m\lambda_{Bragg} = 2n_{eff}\Lambda \qquad (1)$$

where $m$ is the order of diffraction, $\lambda_{Bragg}$ is the central wavelength of the photonic band gap, $n_{eff}$ is the effective refractive index of the photonic crystal, and $\Lambda$ is the pitch, i.e. the length of a unit cell of the structure (in the multilayer case, the thickness of a bilayer). It is evident from the Bragg-Snell law that, to change the position of the photonic band gap, i.e. the structural colour, it is necessary to change either the refractive index of the materials or the size of the unit cell. To change the spectral position of the photonic band gap after the fabrication of the photonic crystal, the unit cell size or the refractive index should be somehow actively changed by an external stimulus.

The possibility to operate by changing the size of unit cell is reported via electroswelling [8] or chemically stimulated swelling [9,10]. The other possibility is to actively change the refractive index of the materials. The most known example is the employment of liquid crystals that show a change of the refractive index induced by the electric field [11–13]. The demonstrated tuning of the optical response of plasmonic materials with the electric field [14] opens to the possibility to employ plasmonic materials in photonic crystals to tune the photonic band gap. An other external stimulus to change the refractive index of materials relates to the use of magneto-optical materials, as for example chalcogenide glasses [15]: with an applied magnetic field and circularly polarized light it is possible to show a shift of the photonic band gap or a defect shift in a microcavity structure.

In this paper we will describe the possibility to tune the photonic band gap of a one-dimensional photonic crystal made with alternated layers of silver and titanium dioxide. We are able to show a shift of about 10 nm with an applied voltage (DC) of 10 V. We will give an explanation of the shift, due to the possibility to change the carrier density of the metal and consequently the plasma frequency; the plasma frequency variation relates to a change of the refractive index, which results in a change of the position of the photonic band gap. We will also describe the shift of a defect in the photonic band gap of a microcavity with the magnetic field since the structure include a layer of magneto-optical material. Finally, we will shortly describe the possibility to tune the gap with UV light when the photonic crystal includes a doped semiconductor, as ITO, that can change the carrier density via UV photodoping.

## 2. METHODS

The silver nanoparticle / $TiO_2$ nanoparticle photonic crystal has been fabricated via spin coating by employing a dispersion of <50 nm size silver nanoparticles in triethylene glycol monoethyl ether (purchased from Sigma Aldrich) and a $TiO_2$ nanoparticle dispersion in water synthetised following the procedure in Ref. [16]. The sample has been annealed for 10 minutes at 350 °C, on a hot plate under fume hood, after each layer deposition.

The microcavity structure employed for the magneto-optical effect is conceived with two $SiO_2/Y_2O_3$ photonic crystals with 10 bilayer and defect of TGG, i.e. $(SiO_2/Y_2O_3)_{10}/TGG/(SiO_2/Y_2O_3)_{10}$. We can write the refractive index of a material impinged with clockwise and counter-clockwise circularly polarized light, and with an applied magnetic field $\vec{B}$ parallel to the propogation of light[17]

$$n_{R,L}(\lambda) = n(\lambda) \pm \frac{V(\lambda)\vec{B}\lambda}{2\pi} \qquad (2)$$

with $\lambda$ the wavelength, $V$ the Verdet constant. For each material we had to take into account the dispersion of the Verdet constant and of the refractive index. For TGG (K4 by Konoshima Chemical) the refractive index is taken from Ref. [18] and the Verdet constant from Ref. [19]. For $SiO_2$ the refractive index is taken from Ref. [20] and the Verdet constant from Refs. [21–23]. For $Y_2O_3$ the refractive index is taken from Ref. [24] and the Verdet constant from Ref. [25]. The obtained reafractive indexes are employed in the transfer matrix method [26–29] in order to simulate the transmission spectrum of the microcavity.

For UV photodoping, to change the spectral position of the photonic band gap, we have fabricated a photonic crystal made of ITO nanoparticles and $SiO_2$ nanoparticles, as described in Ref. [30].

## 3. RESULTS AND DISCUSSION

In Figure 1 we show two ways to apply an external stimulus on a photonic crystal: in Figure 1a the photonic crystal has been fabricated between two ITO substrates, which work as electrodes, in order to apply an external voltage; in Figure 2 the photonic crystal is placed in a region where a magnetic field is applied.

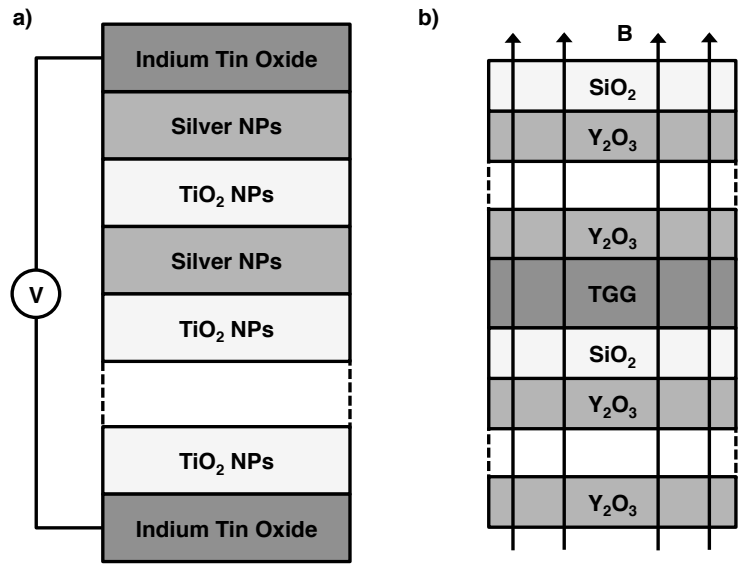

Figure 1. a) Photonic crystal fabricated between ITO substrates in order to apply an external voltage; b) photonic crystal immersed in region of external magnetic field.

In Figure 2a we show the transmission spectrum of the Ag/TiO$_2$ photonic crystal, with the experimental data represented by the dashed curve and a transfer matrix simulation represented by the solid curve. The valley between 450 and 500 nm is ascribed to the silver plasmon while the valley between 600 and 650 nm is ascribed to the photonic band gap. It is evident a mismatch between the simulation and the experiment for the silver plasmon resonance, already discussed in Ref. [1]. Both the transmission valleys shift towards the blue when the external voltage is applied. In Figure 2b we show the shift in energy (eV) of the center of the photonic band gap and it is remarkable that, at about 20 V, the shift starts to saturate. We justify the phenomenon with an increase of the charge carrier density at the silver / TiO$_2$ interface due to the electric field. Such increase in the carrier density increase the energy of the plasma frequency, explaining the blue shift of the plasmon band. Moreover, the increase in the silver carrier density results in a decrease of the dielectric constant and, consequently, the refractive index. This induces a blue shift of the photonic band gap spectral position.

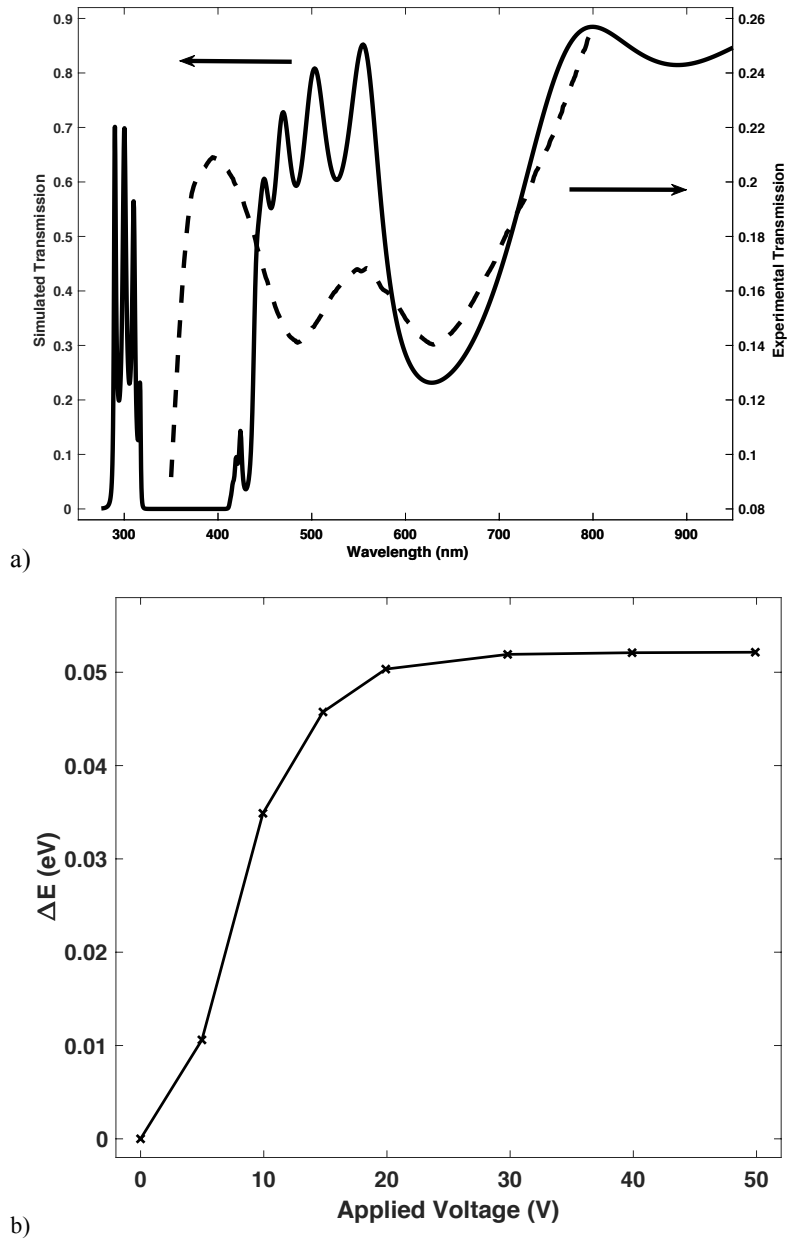

Figure 2. a) Transmission spectrum of the Ag/TiO$_2$ photonic crystal: the dashed curve represents the experimental data, while the solid curve the transfer matrix simulation; b) spectral shift (towards the blue) of the center of the photonic band gap as a function of the applied voltage.

In Figure 3 we show the transmission spectrum for the (SiO$_2$/Y$_2$O$_3$)$_{10}$/TGG/(SiO$_2$/Y$_2$O$_3$)$_{10}$ microcavity under a magnetic field of 5 T, for the clockwise σ$_+$ (dashed curve) and the counter-clockwise σ$_-$ (solid curve) polarizations. The shift of the defect in the photonic band gap is very evident, with value that is more than 20 nm.

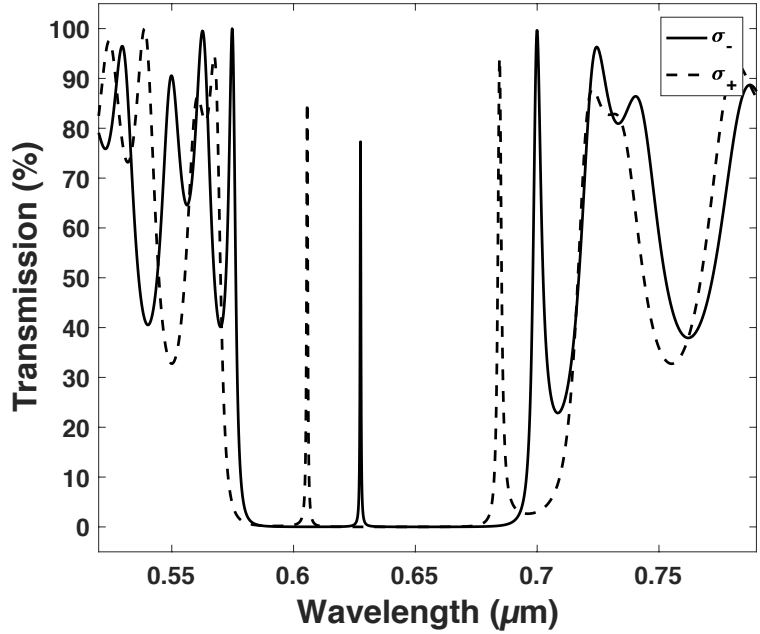

Figure 3. Transmission spectrum of the $(SiO_2/Y_2O_3)_{10}/TGG/(SiO_2/Y_2O_3)_{10}$ microcavity for the clockwise $\sigma_+$ (dashed curve) and the counter-clockwise $\sigma_-$ (solid curve) polarizations, in a magnetic field of 5 T.

Finally, we would like to mention that photonic crystals made with materials that are doped semiconductors as ITO can be photodoped, leading to modification of the photonic band gap via UV photodoping (due to interband transition). The employment of UV light allows also the photonic band gap with ultrafast pulses, leading to ultrafast switchable photonic crystals. In fact, by pumping the photonic crystal at 260 nm (the third harmonic of 150 fs pulse Ti:Sapphire laser), we observed, in the region of the photonic band gap (590 nm), an ultrafast differential transmission signal, with recovery time fo about 20 picoseconds [30].

## 4. CONCLUSION

We have described the possibility to tune the photonic band gap of a one-dimensional photonic crystal made with alternated layers of silver and titanium dioxide. The shift is due to the possibility to change the carrier density of the metal and consequently the plasma frequency, resulting in a variation of the refractive index and a subsequent change of the position of the photonic band gap. We have also described the shift of a defect in the photonic band gap of a microcavity with the magnetic field since the structure include a layer of magneto-optical material. Lastly, we have shortly described the possibility to tune the gap with UV light when the photonic crystal includes a doped semiconductor, as ITO, that can change the carrier density via UV photodoping.

## ACKNOWLEDGEMENT

This project has received funding from the European Union's Horizon 2020 research and innovation programme (MOPTOPus) under the Marie Skłodowska-Curie grant agreement No. [705444], as well as (SONAR) grant agreement no. [734690].